\documentclass[doublecol]{epl2} 
\usepackage{psfrag}

\title{Scaling behavior of self-avoiding walks on percolation clusters}
\shorttitle{Scaling behavior of self-avoiding walks on percolation clusters} 

\author{Viktoriya Blavatska\inst{1,2} \and Wolfhard Janke\inst{1}}
\shortauthor{V. Blavatska 
and W. Janke}

\institute{                    
  \inst{1} Institut f\"ur Theoretische Physik and Centre for Theoretical Sciences (NTZ),\\ Universit\"at Leipzig,
Postfach 100920, 04009 Leipzig, Germany\\
  \inst{2} Institute for Condensed Matter Physics,
National Academy of Sciences of Ukraine, \\
79011 Lviv, Ukraine
}


\abstract{The scaling behavior of self-avoiding walks (SAWs) on the
backbone of percolation clusters in two, three and four dimensions is studied by Monte Carlo simulations.
We apply the pruned-enriched Rosenbluth chain-growth method (PERM). Our numerical results bring about 
the estimates of 
critical exponents, governing the scaling laws 
of disorder averages of the end-to-end distance  of SAW configurations. The effects of finite-size scaling are discussed as well.
}

\begin{document}

\maketitle

\section{Introduction}

The universal configurational 
properties of long, flexible polymer chains in a good solvent are perfectly described by the model of self-avoiding walks (SAWs) on a
regular lattice \cite{desCloizeaux90}. In particular, the average square end-to-end distance 
$\langle R^2\rangle$, and the number of configurations $Z_N$ of SAWs with $N$ steps obey the scaling laws:
\begin{equation}\label{scaling}
 \langle R^2 \rangle
\sim N^{2\nu_{\rm SAW}},\mbox{\hspace{3em}}Z_N \sim z^{N}
N^{\gamma_{\rm SAW}-1},
\end{equation} 
where $\nu_{{\rm SAW}}$ and  $\gamma_{{\rm SAW}}$ are the universal critical exponents that only depend on the
space dimension $d$, and $z$ is a non-universal connectivity constant, depending also on the type of the lattice.  
 The properties of  SAWs on a regular lattice have been studied
in detail both in computer simulations \cite{Rosenbluth55,Madras88,MacDonald92,Li95,Caracciolo98,
MacDonald00} and analytical approaches
\cite{Guillou80,Nienhuis82,Guillou85,Guida98}.  For example, in the
space dimension $d{=}3$ one finds within the frame of the field-theoretical renormalization group
approach  $\nu_{\rm SAW}{=}0.5882\pm 0.0011$ \cite{Guida98} and Monte
Carlo simulations give $\nu_{\rm SAW}{=}0.5877\pm0.0006$ \cite{Li95}. For space dimensions $d$ above the upper
critical dimension $d_{\rm up}{=}4$, the scaling exponent becomes
trivial: $\nu_{\rm SAW}(d\geq4){=}1/2$.

A question of  great interest is how SAWs behave on randomly diluted lattices, which may serve as 
a model of linear polymers in a porous medium. 
Numerous numerical \cite{Kremer81,Lee88,Woo91,Grassberger93,Lee96,Meir89,Lam90,
Nakanishi92,Rintoul94,Ordemann00,Nakanishi91} 
and analytical  \cite{Meir89,Kim83,Barat91,Sahimi84,Rammal84,Kim87,
Roy90,Lam84} studies
lead to the
conclusion that weak quenched disorder, when the
concentration $p$ of lattice sites allowed for the SAWs is higher
than the percolation concentration $p_{c}$, does not influence the scaling of SAWs. 
 The scaling laws (\ref{scaling}) are valid in this case 
with the same exponents, independently of $p$.  More interesting is the case, when 
$p$ equals the critical concentration $p_{c}$ (see Table \ref{dim}) 
and the lattice becomes percolative. 
Studying properties of percolative lattices, one encounters two possible statistical averages. 
In the  first, one considers  only incipient percolation clusters 
whereas  the other statistical ensemble includes all the clusters, which can be found in a percolative lattice. 
For the latter  ensemble of all clusters, the SAW can start on any of the clusters, and for an $N$-step SAW, performed on the $i$th cluster,
 we have $\langle R^2 \rangle \sim l_i^2$, 
where $l_i$ is the averaged  size of the $i$th cluster. 
In what follows, we will be interested in the former case, when SAWs reside only on the percolation cluster.  
In this regime,  the scaling laws (\ref{scaling})
hold with  new exponents $\nu_{p_c}\neq\nu_{{\rm SAW}},\gamma_{p_c}\neq\gamma_{{\rm SAW}} $ 
\cite{Woo91,Grassberger93,Lee96,Rintoul94,Ordemann00,Blavatska04,Janssen07}. A hint
to the physical understanding of this phenomenon is given by the
fact that weak disorder does not change the dimension of a lattice, whereas the percolation cluster itself
is a fractal with fractal dimension $d_{p_c}^F$ dependent on $d$ (see Table \ref{dim}). 
In this way, scaling law exponents of SAWs  change  with the
dimension $d_{p_c}^F$ of the (fractal) lattice on which the walk
resides. Since $d_{\rm up}{=}6$ for percolation
\cite{Stauffer}, the exponent $\nu_{p_c}(d\geq 6){=}1/2$. For the connectivity constant $z_{p_c}$ of SAWs on a percolative lattice 
the  estimate  $z_{p_c}{=}p_c z$ is suggested, where $z$ is the value on the corresponding pure lattice  
\cite{Chakrabarti83}.

\begin{table}[t!]
\small{
\caption{ Critical concentration $p_c$ of site-diluted lattices and fractal
dimensions of percolation cluster $d_{p_c}^F$ and backbone of the percolation cluster $d_{p_c}^B$ 
for different space dimensions $d$.}
\label{dim}
\begin{center}
\begin{tabular}{lccr}
\hline $d$ & $p_c$ & $d_{p_c}^F$ & $d_{p_c}^B$ \\ 
\hline 
2 & $0.592746$ {\cite{Ziff94}}  & $91/49$ {\cite{Havlin87}}  & $1.650\pm0.005$ {\cite{Moukarzel98}}   \\ 
3 & $0.31160$ {\cite{Grassberger92}}  &  $2.51\pm0.02 $ {\cite{Grassberger86}} &  $1.86\pm0.01$ {\cite{Moukarzel98}}  \\ 
4 & $ 0.19688$ {\cite{Paul01}} & $3.05\pm0.05${\cite{Grassberger86}}  &  $1.95\pm 0.05 ${\cite{Moukarzel98}}  \\ 
 \hline 
\end{tabular}
\end{center}
}
\end{table}

Until recently there did not
exist any satisfactory theoretical estimates for scaling law exponents of SAWs on percolation clusters, 
 based on refined field-theoretical approaches. In
particular this was caused by the rather complicated diagrammatic
technique of the  perturbation theory calculations. 
Recently  the field theory
developed by Meir and Harris \cite{Meir89} was reconsidered in
Refs.  \cite{Blavatska04, Janssen07}, where the field theory with complex
interacting fields has been constructed and a special diagrammatic
technique developed. The scaling properties of  a SAW on a
percolation cluster were found to be described by a whole family
of correlation exponents $\nu^{(i)}$, with $\nu^{(1)}{=}\nu_{\rm SAW}$. 

Note that up to date there do also not exist
many studies dedicated to Monte Carlo (MC) simulations of our
problem and they do still exhibit some controversies.
 The first MC study of a SAW statistics on a disordered (diluted) lattice 
in three dimensions was performed in the
work of Kremer \cite{Kremer81}. It indicates no change in the  exponent
 $\nu$ for weak dilution, but for concentrations of dilution
near the percolation threshold a higher value $\nu_{{p_c}}\approx
2/3$ was observed.

 This result was the only
 numerical estimate of $\nu_{p_c}$ for a number of years, until
 Lee {\em et al.} \cite{Lee88,Woo91} performed  MC simulations for a SAW on
 the percolation cluster for square and cubic lattices at dilutions very close to the percolation threshold. Two earlier of these
 works  indicate the rather surprising
 result  that in two dimensions the critical exponent
 $\nu_{{p_c}}$ is {\em not} different compared
 to the pure lattice value. Later, some numerical uncertainties were corrected
 and the value for $\nu_{{p_c}}$ found in two dimensions is in a new universality class.
 This result has been confirmed in a more accurate study of
 Grassberger \cite{Grassberger93}.
 In the case of three and four dimensions, there also exist estimates
 indicating a new universality class \cite{Woo91}, but no
 satisfactory numerical values have been obtained so far.
 It was argued in Ref. \cite{Rintoul94}, that series
 enumerations of all possible SAW
configurations on a percolation cluster give a greater value for $\nu_{p_c}$ (and therefore in better agreement with
theoretical prediction) than that obtained from MC simulations due
to some specific peculiarities of  the latter method.

In the present paper, the so-called chain-growth algorithm
 is applied. Conventional MC methods such as
multicanonical sampling \cite{Berg91} or the Wang-Landau method
\cite{Wang01} expose problems in tackling ``hidden" conformational
barriers in combination with chain update moves which usually
become inefficient at low temperatures, where many attempted moves
are rejected due to the self-avoidance constraint. Rosenbluth chain
growth avoids occupied neighbors at the expense of a
bias. Chain-growth methods with population control such as  PERM (pruned-enriched Rosenbluth method)
\cite{Grassberger97,Hsu03} improve the procedure considerably by utilizing the
counterbalance between Rosenbluth weight and Boltzmann probability.
 PERM has been applied successfully to a wide class of problems, in particular  
to the  $\Theta$-transition of homopolymers \cite{Grassberger97}, trapping of random walkers on absorbing 
lattices\cite{Mehra02}, study of protein folding \cite{Bachmann03}, etc.

\section{Construction of percolation clusters}

\begin{figure}
\begin{center}
\includegraphics[width=4cm]{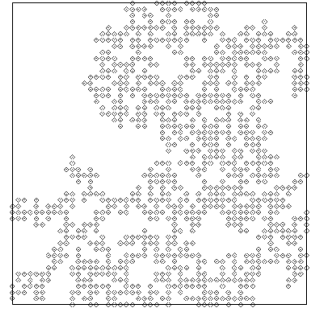}
\caption{Percolation cluster on a $d{=}2$-dimensional regular lattice of edge length $L{=}50$.}
\label{perc}
\end{center}
\end{figure}

We consider site percolation on regular lattices of edge lengths up to $L_{{\rm max}}{=}300,200,50$ in dimensions $d{=}2,3,4$, respectively. Each site of 
the lattice was assigned to be occupied with probability $p_c$ 
(values of critical concentration in different dimensions are given in Table 1), and empty otherwise. 
To extract the percolation cluster, we apply the algorithm of site labeling, based on the one proposed by Hoshen and Kopelman \cite{Hoshen76}. 
If for a given lattice it is not possible to find a cluster that wraps around in all coordinate directions, this disordered lattice is rejected and a new one is constructed.  The typical structure of percolation clusters is presented in Fig. \ref{perc}.  
On finite lattices the definition of spanning clusters is not unique (e.g.,
one could consider clusters connecting only two opposite borders), but
all definitions are characterized by the same fractal dimension and are
thus equally legitimate. The here employed definition has the advantage of
yielding the most isotropic clusters.
Note also that directly at $p=p_c$ more than one spanning cluster can be found in the system, and the probability $P(k)$ for at least $k$ separated clusters  grows with the space dimension as $P(k)\sim \exp(-\alpha k^{ d/(d-1)})$ \cite{Aizenman97,Shchur02}. In our study, we take into account only one cluster per each disordered lattice constructed, in order to avoid presumable correlations of the data.

Since we aim on investigating the scaling of SAWs on a percolative lattice, we are interested rather in the backbone of the percolation cluster,  
which is defined as follows. Assume that each bond (or site) of the cluster is
a resistor and that an external potential drop is applied at two ends of the cluster. The backbone is 
the subset of the cluster consisting of all bonds (or sites) through which the current flows; i.e., it is the structure left when 
all ``dangling ends" are eliminated from the cluster. The SAWs can be trapped in ``dangling ends",  therefore infinitely long chains can 
only exist on the backbone of the cluster. 

\begin{figure}[!t]
\begin{center}
\includegraphics[width=4cm]{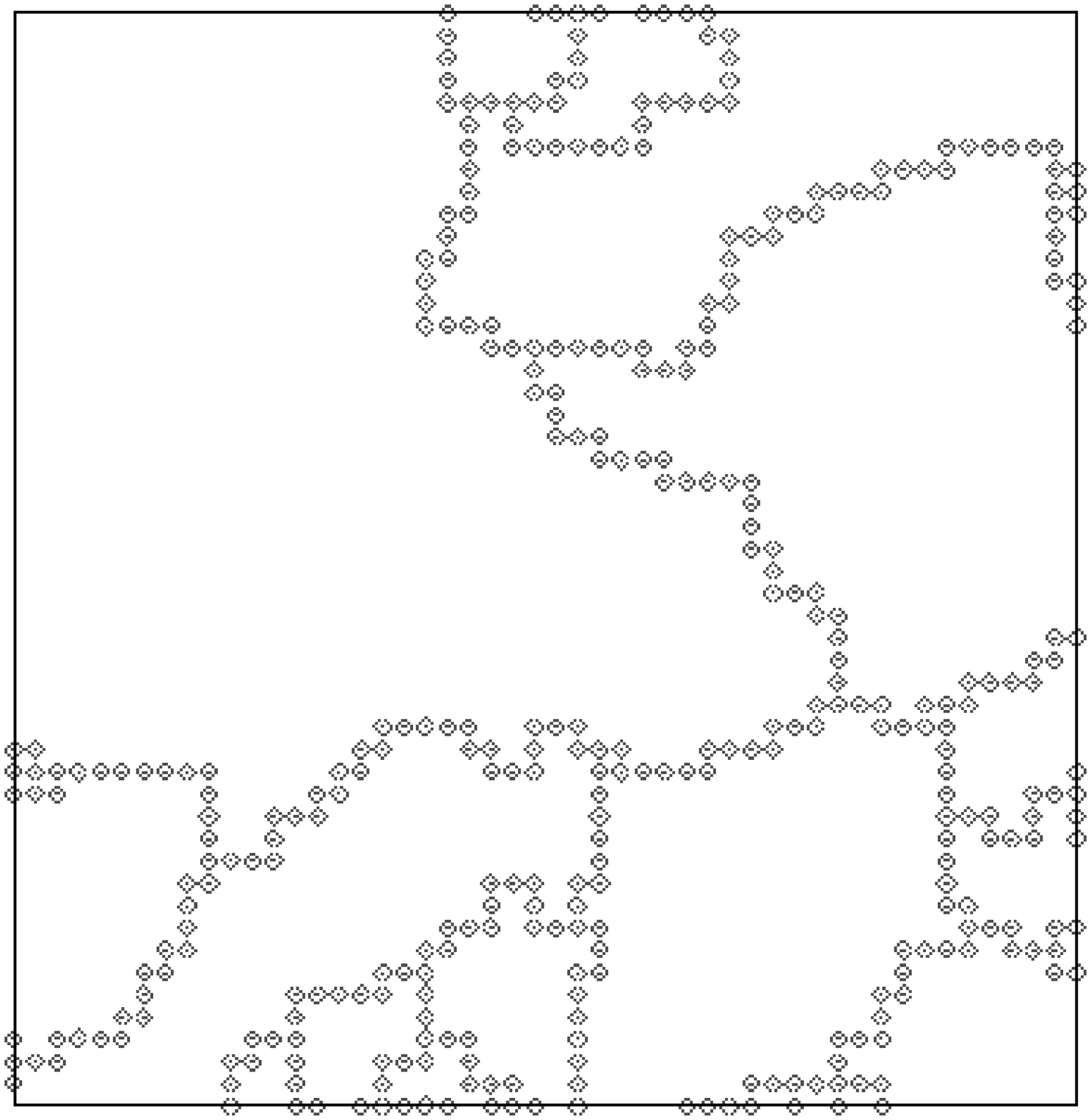}
\hspace*{0.5cm}
\includegraphics[width=4cm]{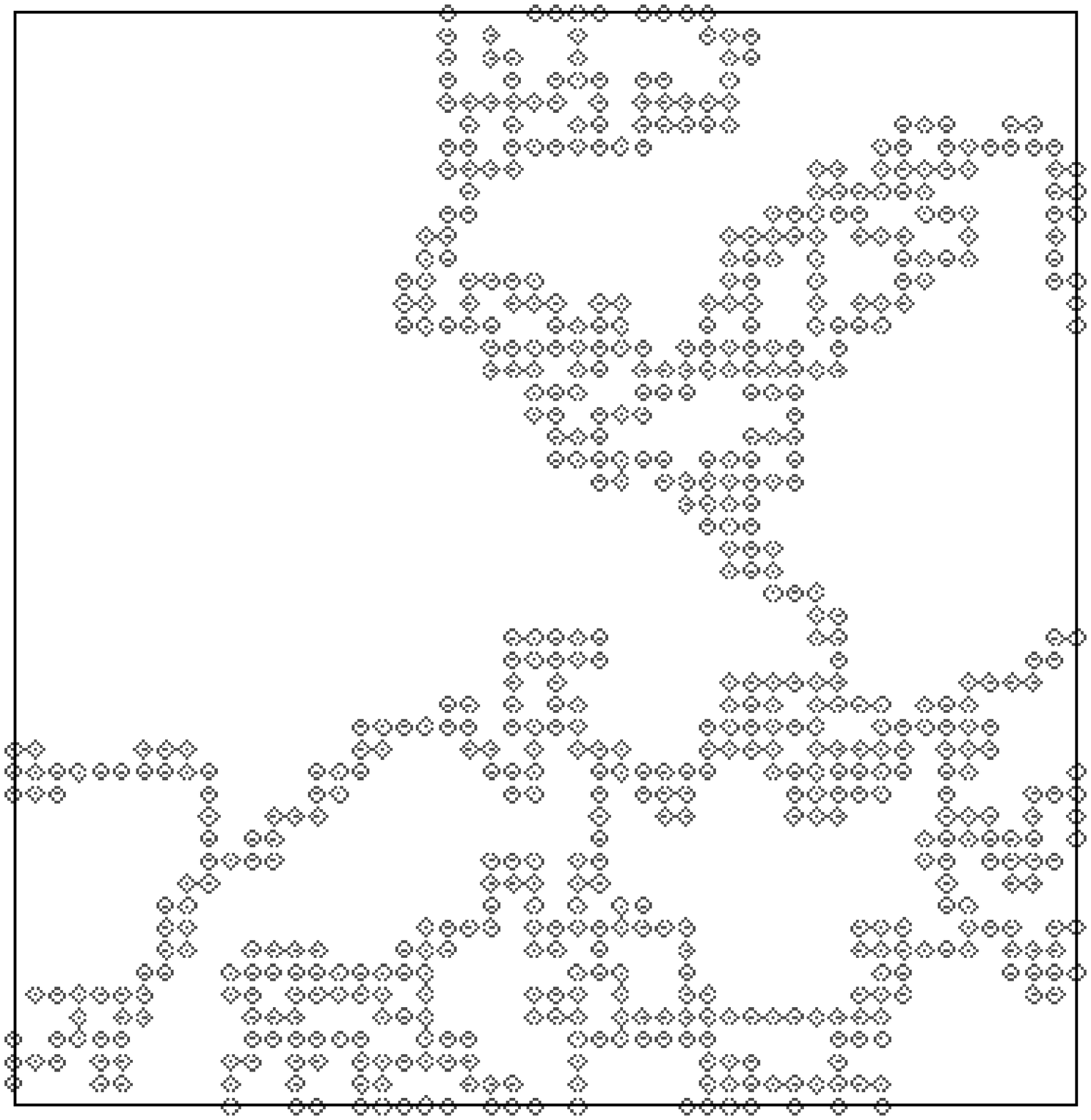}
\caption{Elastic (left) and geometrical (right) backbones of the percolation cluster depicted in Fig. \ref{perc}.}
\label{elas}
\end{center}
\end{figure}

The algorithm for extracting   the backbone of a given percolation cluster was first introduced in \cite{Herrmann84} and improved in \cite{Porto97}. 
This so-called burning algorithm is divided into two parts. First, we choose the starting point --``seed'' -- at the center of the cluster.  
For all the sites on the edge of the lattice,  belonging to the percolation cluster, we find the shortest paths  between the ``seed"  and the given site.  
As a result, we obtain a so-called skeleton or elastic backbone \cite{Havlin84} of the percolation cluster, shown in Fig. \ref{elas}, left.

In the second part of the algorithm, we  consider successively each site of the elastic backbone and check whether a ``loop" starts from this site. 
A ``loop" is a path of sites, belonging to the percolation cluster, which is connected with the elastic backbone by no less than two sites. 
Sites of the elastic backbone together with sites of ``loops" form finally the geometric backbone of the cluster (see Fig. \ref{elas}, right). 

Once a cluster is generated, its fractal dimension in topological (or chemical) space $l$ can be determined  according to \cite{Porto97}:
\begin{equation}
\langle M_B({ l})\rangle\sim{l}^{d_B^{{l}}},
\end{equation}
where $M_B(l)$ is its ``mass" (number of cluster sites) and $ d_B^{{l}}$ is the fractal dimension of the backbone in chemical space. It is related to 
the dimension $d_{p_c}^B$ in coordinate space 
by $d_{p_c}^B{=}d_B^{{l}}d_{{\rm min}}$, where $d_{{\rm min}}$ is the fractal dimension of the shortest path on the backbone and describes the scaling behavior 
between $r$ and ${l}$, i.e. $\langle {l}\rangle \sim r^{d_{{\rm min}}}$, with $d_{{\rm min}}{=}1.130\pm0.004$ in $d{=}2$ \cite{Herrmann88},  $d_{{\rm min}}{=}1.374\pm0.003$ in $d{=}3$ \cite{Grassberger92}, $d_{{\rm min}}{=}1.567$ in $d{=}4$ \cite{Janssen00}. The results for fractal dimensions of the percolation cluster and its 
geometrical backbone in $d{=}2,3,4$ are compiled  in Table 1.

\section{The method}

We use the pruned-enriched Rosenbluth method (PERM), proposed in the work of Grassberger \cite{Grassberger97}.  The algorithm is based on ideas from the very first days of Monte Carlo simulations, the Rosenbluth-Rosenbluth (RR) method \cite{Rosenbluth55} and enrichment strategies \cite{Wall59}. Let us consider the growing polymer chain, 
i.e., the $n$th monomer is placed at a randomly chosen neighbor site of the last placed $(n-1)$th  monomer ($n\leq N$, where $N$ is the total length of the chain). 
In order to obtain correct statistics, if this new site is occupied, any attempt to place a monomer at it results in discarding the entire chain. This leads to an exponential ``attrition", the number of discarded chains grows exponentially with the chain length, which makes the method useless for long chains. 
In the RR method, occupied neighbors are avoided without discarding the chain, but the bias is corrected by means of giving a weight $W_n\sim (\prod_{l{=}2}^n m_l)^{-1}$ 
to each sample configuration at the $n$th step, where $m_l$ is the number of free lattice sites to place the $l$th monomer.
When the chain of total length $N$  is constructed, the new one starts from the same starting point, until the desired number of chain configurations are obtained.  
The configurational averaging for the end-to-end distance $r\equiv \sqrt{R^2(N)}$ is then given by:
\begin{eqnarray}
&&\langle {r} \rangle {=} \frac{\sum_{{\rm conf}} W_N^{{\rm conf}} r^{{\rm conf}}}{\sum_{{\rm conf}} W_N^{{\rm conf}}}{=}\sum_{r}rP(r,N) \label{R},
\end{eqnarray}
where $W_N^{{\rm conf}}$ is the weight of an $N$-monomer chain in a given configuration and $P(r,N)$ is the distribution function for the end-to-end distance. 

\begin{figure}[b!]
\begin{center}
\psfrag{xx}{$r/N^{\nu_{p_c}}$}
\psfrag{yy}{$r\overline{P(r,N)}$}
\includegraphics[width=5.3cm]{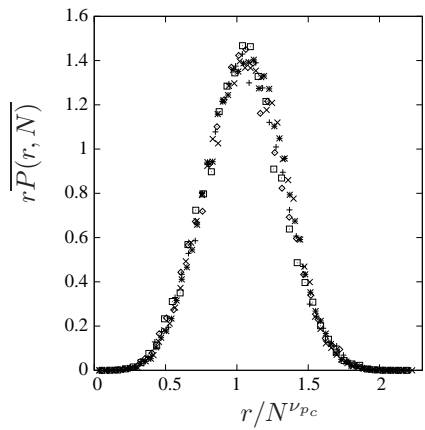}
\caption{Disorder averaged distribution function $r\overline{P(r,N)}$  vs the scaling variable $r/N^{\nu_{p_c}}$ in $d{{=}}3$ dimensions. Lattice size $L{{=}}200$,
number of SAW steps $N{{=}}40$ (squares), $N{{=}}50$ (pluses), $N{{=}}60$ (diamonds), $N{{=}}70$ (crosses), $N{{=}}80$ (stars).}
\label{pr}
\end{center}
\end{figure}
\begin{figure}[t!]
\begin{center}
\psfrag{xx}{$r/N^{\nu_{p_c}}$}
\psfrag{yy}{$r\overline{P(r,N)}$}
\includegraphics[width=5.3cm]{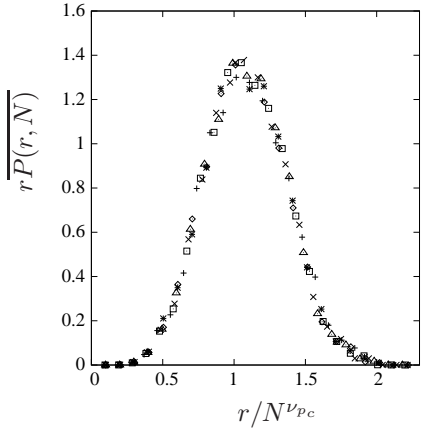}
\caption{Disorder averaged distribution function $r\overline{P(r,N)}$  vs the scaling variable $r/N^{\nu_{p_c}}$ in $d{{=}}4$ dimensions. Lattice size $L{{=}}50$,
number of SAW steps $N{{=}}15$ (squares), $N{{=}}18$ (triangles), $N{{=}}20$ (pluses), $N{{=}}25$ (crosses), $N=30$ (stars).}
\label{pr2}
\end{center}
\end{figure}

While the chain grows by adding monomers, its weight will fluctuate. PERM suppresses these fluctuations by pruning configurations with too small weights, and by enriching the sample with copies of high-weight configurations \cite{Grassberger97}. These copies are made while the chain is growing, and continue to grow independently of each other. Pruning and enrichment are performed by choosing thresholds $W_n^{<}$
and $W_n^{>}$ depending on the estimate of the partition sums of the $n$-monomer chain. These thresholds are continuously updated as the simulation progresses. The zeroes iteration is a pure chain-growth algorithm without reweighting. After the first chain of full length has been obtained, we switch to $W_n^{<}$,
 $W_n^{>}$. If the current weight $W_n$ of an $n$-monomer chain is less than $W_n^{<}$, a random number $r{=}{0,1}$ is chosen; if $r{=}0$, the chain is discarded, otherwise it is kept and its weight is doubled. Thus, low-weight chains are pruned with probability $1/2$. If $W_n$ exceeds  
$W_n^{>}$, the configuration is doubled and the weight of each copy is taken as half the original weight. 
For updating the threshold values we apply similar rules as in \cite{Hsu03,Bachmann03}: $W_n^{>}{=}C(Z_n/Z_1)(c_n/c_1)^2$ and $W_n^{<}{=}0.2W_n^{>}$, where $c_n$ denotes the number of created chains having length $n$, and the parameter $C$ controls the pruning-enrichment statistics. 
After a certain number of chains of total length $N$ is produced, the iteration is finished and a new tour starts. 
We adjust the pruning-enrichment control parameter such that on average 10 chains of total length $N$ are generated per each iteration \cite{Bachmann03}, and perform $10^6$ iterations. 
Also, what is even more important for efficiency, in almost all iterations at least one such a chain was created. 
\begin{figure}[!b]
\begin{center}
\psfrag{xx}{$\ln N$}
\psfrag{yy}{$\ln {\overline{\langle r\rangle}}$}
\includegraphics[width=6cm]{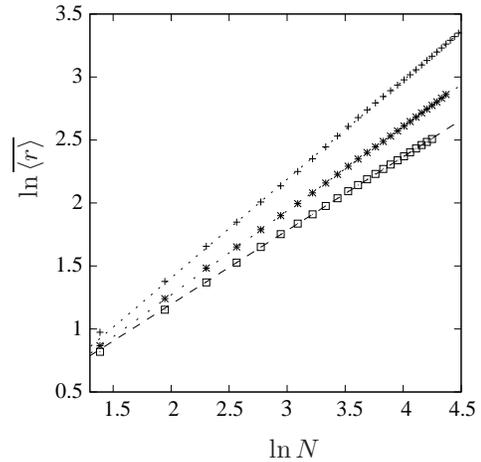}
\caption{Disorder averaged  end-to-end distance vs number of steps in double logarithmic scale for SAWs on the backbone of percolation 
clusters in 
 $d{=}2$ (pluses), $d{=}3$ (stars), $d{=}4$ (squares).
Lines represent linear fitting, statistical error bars are of the size of symbols.}
\label{sawr}
\end{center}
\end{figure}

 \section{Results} 

To study scaling properties of SAWs on the backbone of percolation clusters, we choose as the starting point  the ``seed" of the cluster, and apply the  
PERM algorithm, taking into account, that a SAW can have its steps only on the sites belonging to the backbone of the percolation cluster. 
In the given problem,  we have to perform two types of averaging: the first average is performed over all SAW configurations on a single backbone according to (\ref{R});
the second average is carried out over different realizations of disorder, i.e. over many backbone configurations:
\begin{eqnarray}
&&\overline{\langle r \rangle}{=}\frac{1}{C}\sum_{c{=}1}^C \langle r\rangle_c{=}\sum_{r}r {\overline {P(r,N)}},\label{rav}\\
&&\overline {P(r,N)}{=}\frac{1}{C}\sum_{c{=}1}^C P_c(r,N). \label{avprob} 
\end{eqnarray}
Here, $C$ is the number of different clusters, the index $c$ means that a given quantity is calculated on the cluster $c$, 
$\overline{P(r,N)}$ is the distribution function,  averaged over cluster configurations.
  
\begin{table}[t!]
\caption{Results  of linear fitting of obtained results for $\overline{\langle r\rangle}$ 
for SAWs in $d{=}3$ dimensions on the backbone of percolation clusters, $L{=}200$. $\chi^2$ denotes the sum of squares of normalized deviation from the regression line, 
$DF$ is the number of degrees of freedom.}
\label{3d}
\begin{center}
\begin{tabular}{rccr}
\hline
$N_{{\rm min}}$ & $\nu_{p_c}$ & $a$ & $\chi^2/DF$ \\
\hline
6 & 0.665         $\pm$ 0.003 &  0.946         $\pm$ 0.003 & 2.783 \\ 
11 &  0.668         $\pm$ 0.003 & 0.935         $\pm$ 0.004 & 2.269\\
16 &  0.669         $\pm$ 0.003 & 0.930         $\pm$ 0.004 & 2.054\\
21 &  0.669         $\pm$ 0.003 & 0.924         $\pm$ 0.004 &  1.345\\
26 &   0.667         $\pm$ 0.002   & 0.930         $\pm$ 0.006 & 0.743\\
31 & 0.668         $\pm$ 0.002 &  0.934         $\pm$ 0.008 & 0.844\\
\hline
\end{tabular}
\end{center}
\end{table}

The case of so-called ``quenched disorder" is considered, where 
the average over different realizations of disorder is taken after the configurational average has been performed. 
As it was pointed out in \cite{Grassberger93},  
the correctness of results, obtained in the picture of ``quenched" disorder, depends on whether the location of the starting point of a SAW  is fixed 
while the configurational averaging is performed, or not. In the latter case, one has to average over all locations and effectively this corresponds to 
the case of annealed disorder. Thus, as we have already stated above, we start each configuration of a SAW
on the same site -- the ``seed" of the backbone of a given percolation cluster. 
We use lattices of the size up to $L_{{\rm max}}{=}300, 200, 50$ in $d{=}2,3,4$, respectively, and performed averages over 
1000 percolation clusters in each case. 

The disorder averaged distribution function (\ref{avprob}) can be written in terms of the scaled variables 
$r/\overline{\langle r\rangle}$ as:
\begin{equation}
r\overline{P(r,N)}\sim f(r/\overline{\langle r\rangle})\sim f(r/N^{\nu_{p_c}}).
\end{equation}
The distribution function is normalized according to $\sum_{r}{\overline {P(r,N)}}{{=}}1$. The numerical results for the distribution function in $d{{=}}3$ and $d{{=}}4$ are 
shown in Figs. \ref{pr} and \ref{pr2} for different $N$. When plotted against the scaling variable $r/N^{\nu_{p_c}}$, the data are indeed found to nicely collapse onto a single curve, using our 
values for the  exponent $\nu_{p_c}$ reported in Table 3 below.


To estimate the critical exponents $\nu_{p_c}$, linear least-square fits with lower cutoff for the number of steps $N_{{\rm min}}$ are used.
The $\chi^2$ value (sum of squares of normalized deviation from the regression line)  serves as a test of the goodness of fit (see Fig. \ref{sawr} and Table \ref{3d}).

\begin{figure}[t!]
\hspace*{0.1cm}
\psfrag{xx}{$\ln N$}
\psfrag{yy}{$\ln \langle r\rangle$}
\includegraphics[width=4.0cm]{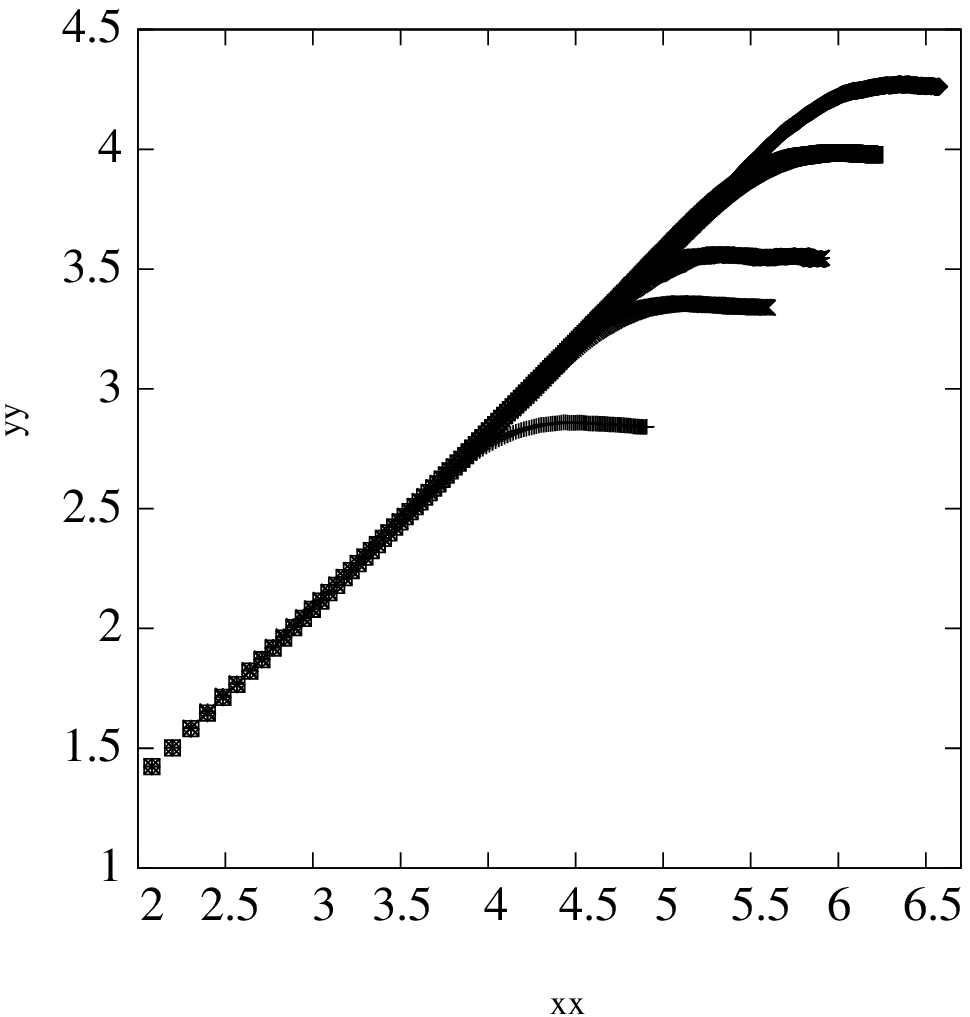}
\hspace*{0.2cm}
\psfrag{xx}{$\ln N$}
\psfrag{yy}{$\ln {\overline{\langle r \rangle}}$}
\includegraphics[width=4.0cm]{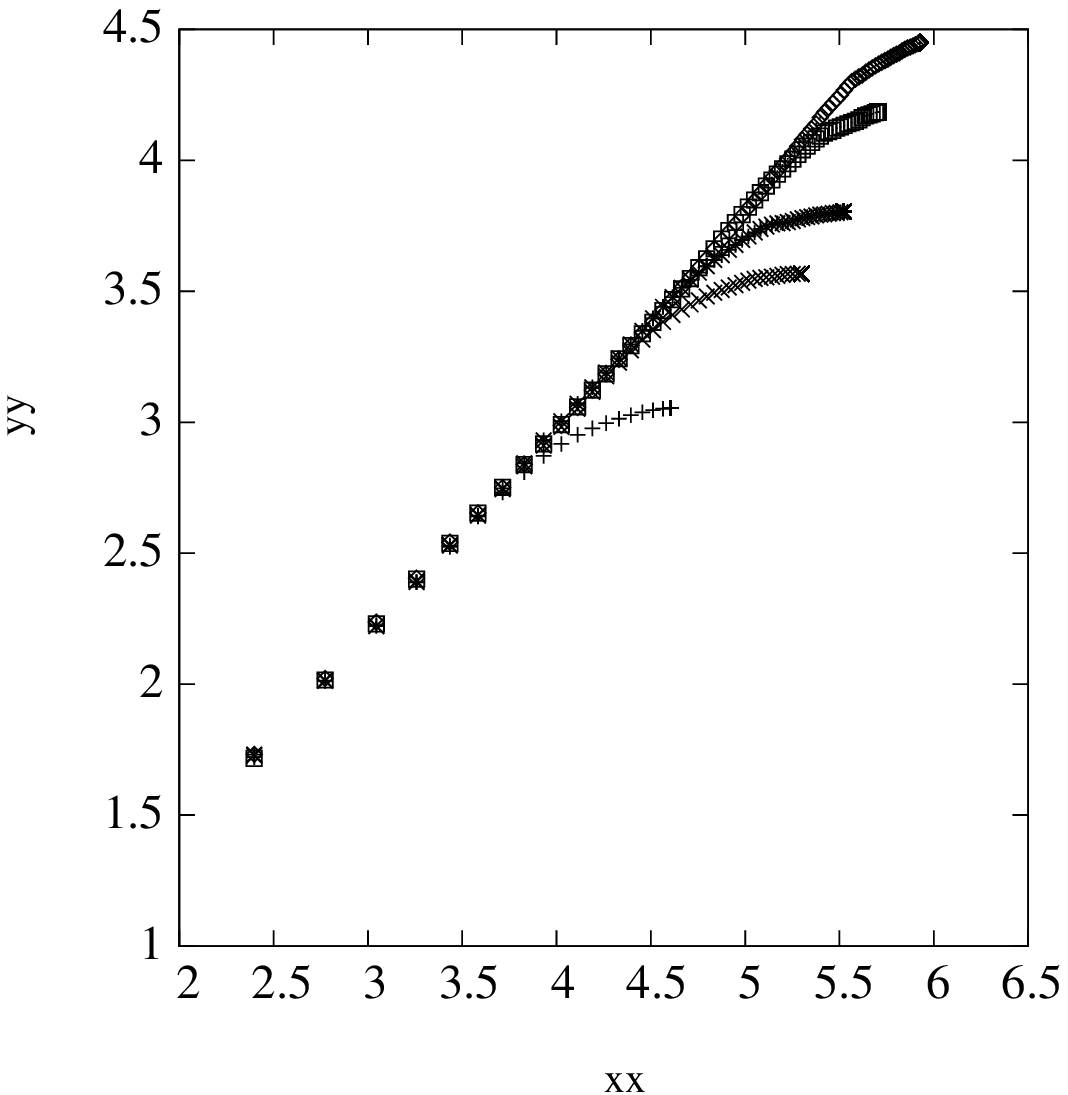}
\caption{ 
Averaged end-to-end distance vs number of steps on a double logarithmic scale for SAWs on a pure lattice (left) 
and on the backbone of a percolation cluster (right) in $d{{=}}2$. In both cases the lattice size $L$ changes from below: $L{{=}}50, 80, 100, 150, 200$.
 Error bars are of the size of symbols.}
\label{finit2}
 \end{figure}

Since  we can construct lattices only of a finite size $L$, it is not possible to perform very long SAWs on it. 
For each $L$, the scaling (\ref{scaling}) holds  only up to some ``marginal" number of SAWs steps
$N_{{\rm marg}}$, as it is shown in Fig. \ref{finit2}. We take this into account when analyzing the data obtained; 
for each lattice size we are interested only in values of $N<N_{{\rm marg}}$, which results in effects of finite-size scaling for critical exponents.   

\begin{figure}[b!]
\hspace*{0.15cm}
\psfrag{xx}{$N/L^{\omega}$}
\psfrag{yy}{$g(N/L^{\omega})$}
\includegraphics[width=4.0cm]{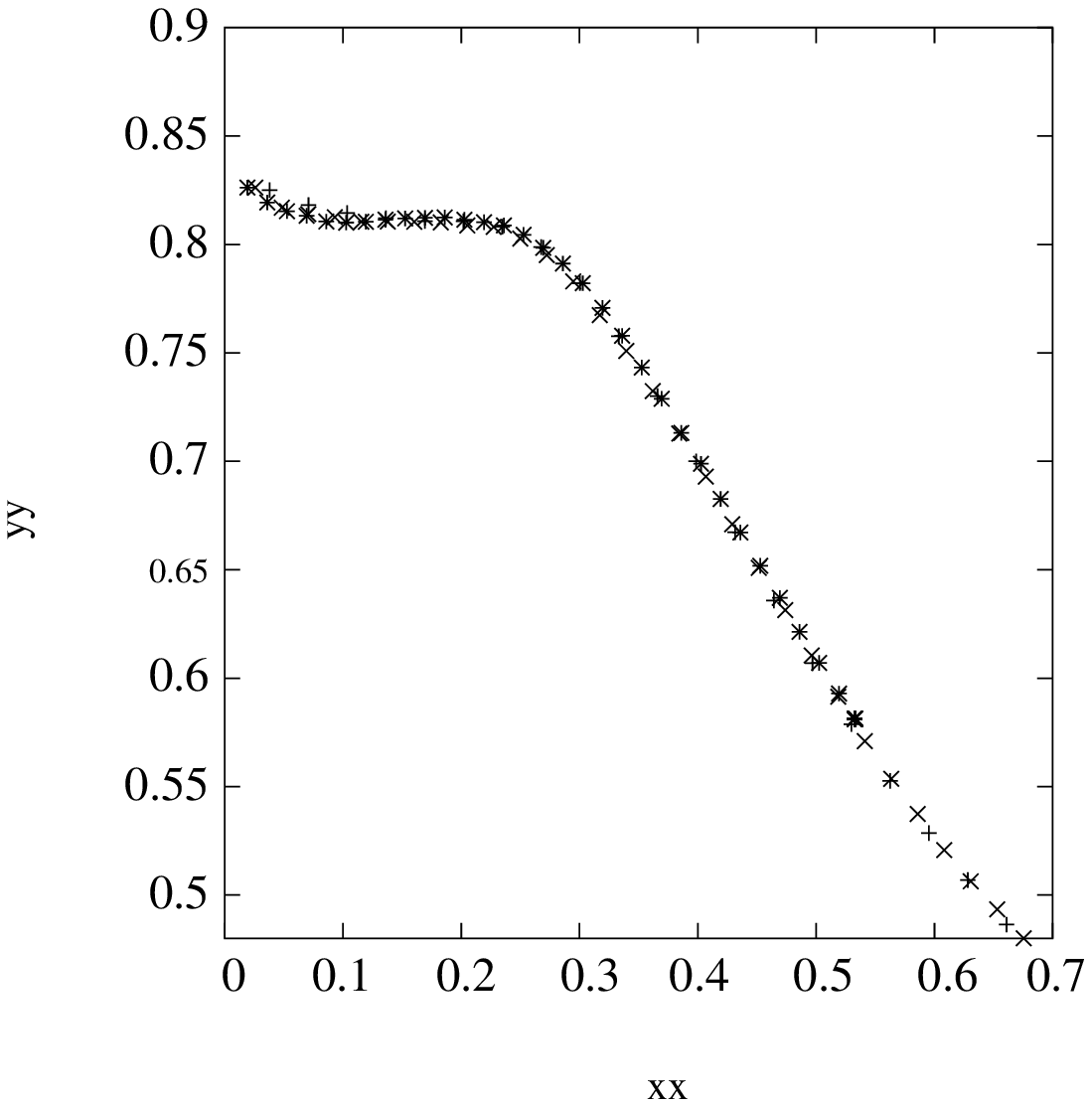}
\hspace*{0.2cm}
\psfrag{xx}{$N/L^{\omega}$}
\psfrag{yy}{$g(N/L^{\omega})$}
\includegraphics[width=4.0cm]{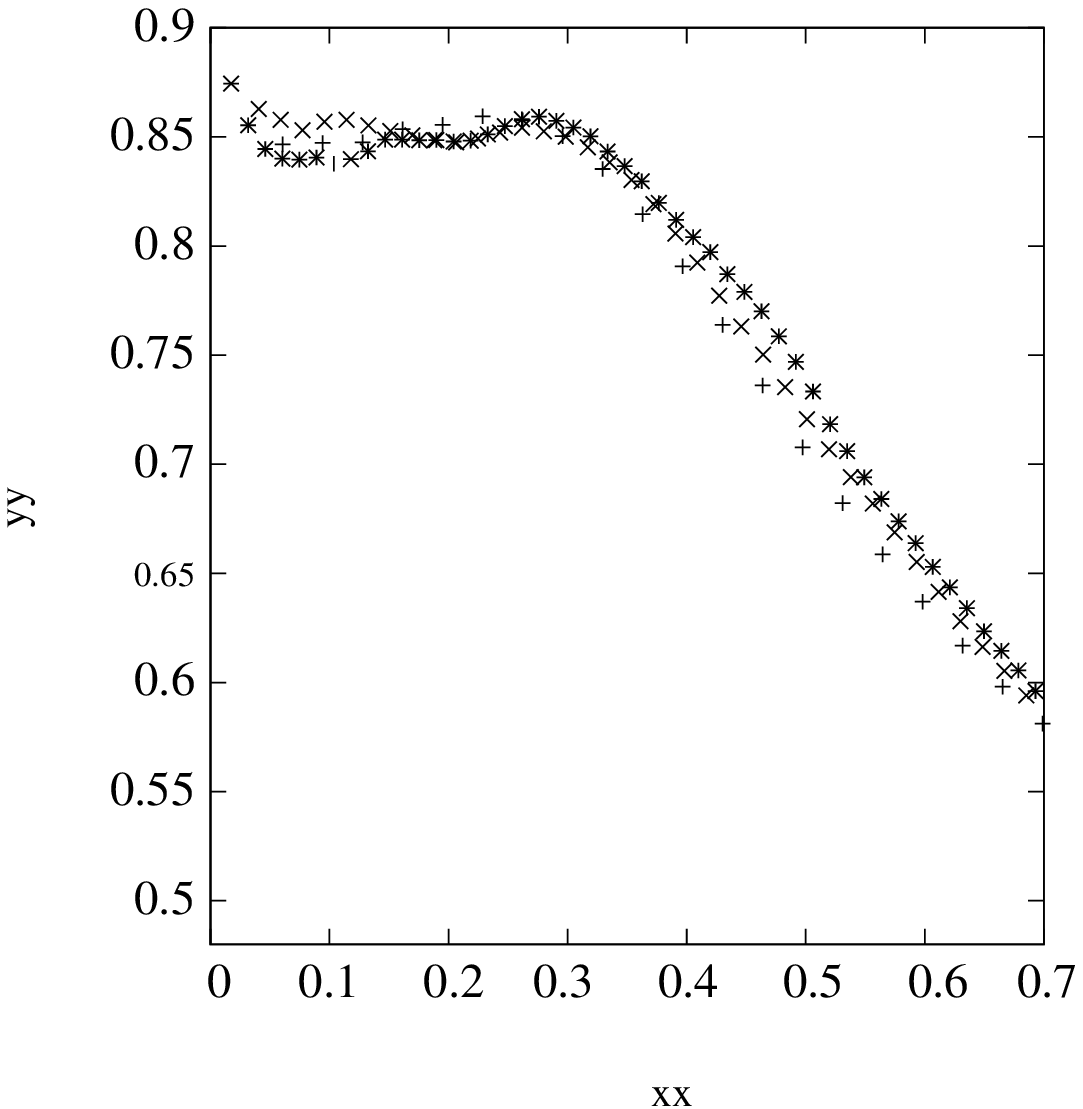}
\caption{ 
 The scaling function $g(N/L^{\omega})$ as a function of its argument at data collapse for  three different lattice sizes $L{{=}}100,150,200$ 
in $d{{=}}2$. Left: pure lattice, right: backbone of the percolation cluster.}
\label{scal2}
\end{figure}

Let us assume that $N_{{\rm marg}} \sim L^{\omega}$,
and for a SAW confined inside a lattice with size $L$ finite-size scaling holds:  
\begin{equation}
\langle r \rangle  \sim N^{\nu}g(\frac{N}{L^{\omega}}),
\label{scal}
\end{equation}
where $$g(x){=}{\rm const}  {\mbox\,\,\, {\rm when}\,\,\,} N \ll L^{\omega}$$
  so that Eq. (\ref{scaling}) is recovered. 
The crossover occurs at ${\langle r\rangle}\sim L,  N{=}N_{{\rm marg}}$, which leads to $\omega{=}1/\nu$. 
Here, $\nu$ stands for $\nu_{{\rm SAW}}$ or $\nu_{p_c}$ for the cases of 
the pure lattice and backbone of percolation cluster, respectively. 
Similar scaling properties have already been observed in problems of random walks in confined environment in Ref. \cite{Reis95}. 

\begin{table}[!t]
{\small
\caption{ \label{allnu} The exponent $\nu_{p_c}$ for a SAW on a
percolation cluster. FL: Flory-like theories, EE: exact
enumerations, RS, RG: real-space and field-theoretic RG. For SAWs on the regular lattice one has:
 $\nu_{{\rm SAW}}(d{{=}}2){=}3/4$~\cite{Nienhuis82}, $\nu_{{\rm SAW}}(d{{=}}3){=}0.5882(11)$~\cite{Guida98},
  $\nu_{{\rm SAW}}(d\geq 4){=}1/2$.}
\begin{tabular}{|r| c| c|  c|  }
\hline $\nu_{p_c} \setminus d$  & 2 & 3 & 4 \\ 
  \hline  FL \cite{Roy90} & 0.77& 0.66 & 0.62 \\ 
 \hline  EE  \cite{Rintoul94}&0.770(5)& 0.660(5)&\\
\cite{Ordemann00}&0.778(15)& 0.66(1)& \\
\cite{Ordemann00}&0.787(10)& 0.662(6)&\\
 \hline  RS
\cite{Lam84} & 0.767 & &  \\ \cite{Sahimi84} & 0.778&0.724
&\\ \hline RG  \cite{Blavatska04} & 0.785 & 0.678& 0.595 \\
\cite{Janssen07} & 0.796 & 0.669& 0.587 \\
\hline MC 
\cite{Woo91}
  & 0.77(1)  & &   \\
\cite{Grassberger93} & 0.783(3) & & 
\\
\cite{Lee96} & & 0.62--0.63 &0.56--0.57 
\\
our results  & $ 0.782\pm 0.003$ & $0.667\pm 0.003$ & $0.586\pm 0.003$ 
\\
 \hline
\end{tabular}
}

\end{table}

Having estimated values for the critical exponent $\nu_{p_c}$, presented in Table 3, we can proceed with 
testing the finite-size scaling assumption~(\ref{scal}). 
When plotted against the scaling variable $N/L^{\omega}$, the data for 
different lattice sizes $L$ should collapse  onto a single curve if we 
have found the correct values for the critical exponents. The numerical results for the scaling function $g(N/L^{\omega})$ both 
for the pure lattice (for comparison)
 and the backbone of percolation clusters are presented in  Fig. \ref{scal2}. Note, that our estimation of the exponent $\nu_{{\rm SAW}}$ in 
two dimensions gives $0.745\pm0.002$.

\section{Conclusions} 
The present paper concerns the universal configurational properties of SAWs on 
 percolative lattices. 
The statistical averaging  was performed  on the backbone  of the incipient
percolation cluster, which has a fractal structure and is characterised by fractal dimension $d_{p_c}^B$.
 Note, that up to date there do not exist
many works dedicated to Monte Carlo (MC) simulations of our
problem and they do still exhibit some controversies.
 In particular, in the case of four dimensions, there exist only estimates,
 indicating a new universality class \cite{Woo91}, but no
 satisfactory numerical values for critical exponents have been obtained so far.

Applying the pruned-enriched Rosenbluth method (PERM), we studied SAWs on the backbone of percolation clusters, 
using lattices of size up to $L_{{\rm max}}{=}300, 200, 50$ in $d{=}2,3,4$, respectively, and performing averages over 
1000 clusters in each case. Our results bring about numerical values of critical exponents,
governing the end-to-end distance of SAWs in a new universality class. The effects of finite lattice size are discussed as well.

\acknowledgments V.B. is grateful for support through the ``Marie Curie International Incoming Fellowship" EU Programme  and
to the Institut f\"ur Theoretische Physik, Universit\"at Leipzig,
for hospitality.

\end{document}